\documentstyle[12pt,epsf]{article}

\begin{document}

\begin{titlepage}

\begin{flushright}
MIT-CTP-3140\\
\end{flushright}

\vskip 2.5cm

\begin{center}
{\Large \bf Photon Decay at the Schwarzschild Horizon}
\end{center}

\vspace{1ex}

\begin{center}
{\large B. Altschul\footnote{Department of Mathematics, {\tt baltschu@mit.edu}}
and R. Jackiw\footnote{Center for Theoretical Physics, {\tt
jackiw@mitlns.mit.edu}}}

\vspace{5mm}
{\sl Massachusetts Institute of Technology} \\
{\sl Cambridge, MA, 02139-4307 USA} \\

\end{center}

\vspace{2.5ex}

\medskip

\centerline {\bf Abstract}

\bigskip

A recent proposal that gravity theory is an emergent phenomenon also entails
the possibility of photon decay near the Schwarzschild event horizon. We
present a possible mechanism for such decay, which utilizes a dimensional
reduction near the horizon.

\bigskip 

\end{titlepage}

\newpage

In a recent paper~\cite{ref-chapline}, the question is raised whether
classical General Relativity can exist as an emergent phenomenon---as the
low-energy limit of an underlying quantum system. In this view, the singularity
at the Schwarzschild event horizon represents a failure of the effective
description owing to the divergence of a characteristic coherence length.  We
shall introduce an additional element to this model: a natural change in the
dimensionality of virtual particle loop integrals of the quantum system near
the horizon.  This change will have important implications. In particular, it
may cause photons to decay when they near the event horizon of a black hole, as
suggested in~\cite{ref-chapline}.

In the Schwarzschild metric, the line element is
\begin{equation}
ds^{2}=\left(1-\frac{r_{S}}{r}\right)dt^{2}-\left(1-\frac{r_{S}}{r}
\right)^{-1}dr^{2}-r^{2}(d\theta^{2}+\sin^{2}\theta\, d\phi^{2}).
\end{equation}
(The velocity of light is set to one.) Classically, this metric concentrates
the motion in the radial direction near $r=r_{S}$. This can be seen by
examining the spatial line element
\begin{equation}
dl^{2}=\left(1-\frac{r_{S}}{r}\right)^{-1}\left[dr^{2}+\left(1-\frac{r_{S}}{r}
\right)r^{2}d\Omega^{2}\right].
\end{equation}
For $r$ near $r_{S}$, the angular variables are suppressed, and motion is
confined to the two-dimensional $t$-$r$ subspace.

More specifically, consider the geodesic equation, $\ddot{x}^{\mu}+\Gamma^{\mu}
_{\nu\rho}\dot{x}^{\nu}\dot{x}^{\rho}=0$, where $\dot{x}^{\mu}$ is the
derivative of the position with respect to an affine parameter.

The temporal and angular equations,
\begin{eqnarray}
\ddot{t}+\frac{r_{S}}{r(r-r_{S})}\dot{t}\dot{r} & = & 0,\\
\ddot{\theta}+2\frac{\dot{r}}{r}\dot{\theta}-\frac{1}{2}\sin 2\theta\,\dot
{\phi}^{2} & = & 0,\\
\ddot{\phi}+\frac{2}{r}\dot{r}\dot{\phi}+2\cot\theta\,\dot{\theta}\dot{\phi} &
= & 0,
\end{eqnarray}
can be readily integrated to give
\begin{eqnarray}
\label{eq-timeODE}
\dot{t}\frac{r-r_{S}}{r} & = & \tau,\\
\label{eq-phiODE}
r^{2}\sin^{2}\theta\,\dot{\phi} & = & m,\\
(r^{2}\dot{\theta})^{2}+\frac{m^{2}}{\sin^{2}\theta} & = & l^{2}.
\end{eqnarray}
$\tau$, $m$, and $l^{2}$ are integration constants. Near the Schwarzschild
horizon we may allow all particles to be massless and take the geodesics to be
null. Then the constants will diverge, but we retain their ratios, which
remain finite. From the $\theta$ equation, it is clear that for $\theta=\frac
{\pi}{2}$ and $\dot{\theta}=0$, $\ddot{\theta}=0$ also, so the motion remains
in the equatorial plane, with $m^{2}=l^{2}$.

Using the other three equations, the radial equation
\begin{equation}
\ddot{r}+\left(1-\frac{r_{S}}{r}\right)\frac{r_{S}}{2r^{2}}\dot{t}^{2}+\left(1-
\frac{r_{S}}{r}\right)^{-1}\frac{r_{S}}{2r^{2}}\dot{r}^{2}
+(r_{S}-r)(\dot{\theta}^{2}+\sin^{2}\theta\,\dot{\phi}^{2})=0,
\end{equation}
becomes
\begin{equation}
\ddot{r}+\frac{\tau^{2}r_{S}}{2r(r-r_{S})}-\frac{r_{S}}{2r(r-r_{S})}\dot{r}^{2}
-\frac{l^{2}(r-r_{S})}{r^{4}}=0.
\end{equation}
In this form, the equation may be integrated to give
\begin{equation}
\label{eq-radial}
\dot{r}=\pm\tau\sqrt{1-\left(1-\frac{r_{S}}{r}\right)\frac{\lambda^{2}}{r
^{2}}},
\end{equation}
where $\lambda=l/\tau$ and a possible integration constant is set to zero for
null geodesics.

Dividing the $\phi$ and $r$ equations,~(\ref{eq-phiODE}) and~(\ref{eq-radial}),
by the $t$ equation~(\ref{eq-timeODE}), gives at $\theta=\frac{\pi}{2}$, $m=l$,
\begin{eqnarray}
\label{eq-rtime}
\frac{d\phi}{dt} & = & \left(1-\frac{r_{S}}{r}\right)\frac{\lambda}{r^{2}},\\
\label{eq-phitime}
\frac{dr}{dt} & = & \pm\left(1-\frac{r_{S}}{r}\right)\sqrt{1-\left(1-\frac
{r_{S}}{r}\right)\frac{\lambda^{2}}{r^{2}}}.
\end{eqnarray}

Consider a photon trajectory just outside the event horizon, at radius $r=r_{S}
+\Delta$. For $\Delta\ll r_{S}$, the time-evolution equations~(\ref{eq-rtime})
and~(\ref{eq-phitime}) become
\begin{eqnarray}
\label{eq-DeltaODE}
\frac{d\Delta}{dt} & = & \pm\frac{\Delta}{r_{S}}\sqrt{1-\frac{\lambda^{2}}{r
_{S}^{3}}\Delta},\\
\frac{d\phi}{dt} & = & \frac{\lambda}{r_{S}^{3}}\Delta.
\end{eqnarray}
The equations, with the lower sign in~(\ref{eq-DeltaODE}), may be integrated,
yielding
\begin{eqnarray}
\label{eq-Delta}
\Delta & = & \frac{r_{S}^{3}/\lambda^{2}}{\cosh^{2}\frac{t}{2r_{S}}}\\
\label{eq-phi}
\phi & = & \frac{2r_{S}}{\lambda}\tanh\frac{t}{2r_{S}}.
\end{eqnarray}
Equations~(\ref{eq-Delta}) and~(\ref{eq-phi}) show that if $\frac{\lambda}{r
_{S}}$ is sufficiently small ($\ll\pi$), then the radial motion will be much
more rapid than the angular motion. This is the regime we are interested in,
where the concentration of motion in the radial direction reduces some aspects
of the system to 1+1 dimensions.

We shall consider a much more profound reduction in the dimensionality of the
system. Taking local coordinates $r$, $x$, and $y$, the spatial line element is
\begin{equation}
dl^{2}=-g_{rr}dr^{2}+dx^{2}+dy^{2}.
\end{equation}
In the model~\cite{ref-chapline}, the singularity at $r=r_{S}$ represents a
real physical effect, not merely a coordinate artifact, so these are very
natural linear coordinates. If we suppose that $p^{r}$, $p^{x}$, and $p^{y}$
are cut off at the same scale in a loop integral, ${\bf p}^{2}\equiv-p_{i}p
^{i}$ will be dominated by $-g_{rr}(p^{r})^{2}$. (This supposition strongly
breaks general covariance, of course.) This is the kind of situation we would
like to analyze. However, a sharp momentum cutoff is not gauge invariant; to
study the photon self-energy, we shall translate this idea into gauge-invariant
language, using dimensional regularization.

In the renormalization prescription described above, one of the spatial
dimensions provides the dominant contribution to ${\bf p}^{2}$. In the language
of dimensional regularization, this can be seen as a reduction in the
effective dimensionality $d$ of the momentum integral to $d<4$.

We shall find the effective dimensionality by examining the volume element of
this system, because the momentum cutoff in a given direction and the volume
contribution of that direction are closely related. To see this, consider for
the moment a theory that is regulated by a lattice at short distances. The
volume of a region counts the number of lattice points in that region. The
lattice spacing $a_{i}$ in a given direction governs the density of lattice
points along that axis, so a length $L$ in the $x_{i}$-direction contributes an
amount $\alpha\frac{L}{a_{i}}$ to the volume, where $\alpha$ is a scaling
constant independent of direction. The lattice spacing also corresponds
directly to the momentum cutoff in that direction, $p_{i}^{max}=\frac{\pi}{a
_{i}}$. So the dependence of the volume on a given dimension and the momentum
cutoff in that direction are intimately linked.

So we look for an expression for the effective dimensionality (to be used in
dimensional regularization) in terms of the volume element. In Schwarzschild
space-time, the spatial volume element is
\begin{equation}
dV=\sqrt{g}\, dr\, d\theta\, d\phi.
\end{equation}
Here, $g=-g_{rr}r^{4}\sin^{2}\theta$ is the determinant of the spatial metric
$-g_{ij}$. The classical geodesic problem suggests that the radial direction
should always contribute one effective dimension, while the angular directions
may contribute less than one. We wish to determine the effective dimensionality
by an integral over an effective volume. To a achieve a reasonable result, the
effective volume is defined by rescaling a factor of $\sqrt{-g_{rr}}$ from each
direction. So the new volume element reads
\begin{equation}
dV'=\frac{r^{2}\sin\theta}{-g_{rr}}\, dr\, d\theta\, d\phi=g_{tt}dV_{E},
\end{equation}
where $dV_{E}$ is a Euclidean volume element.

From $dV'$, we need to find an expression for the number of effective
dimensions. This expression should have several properties. The dimension
corresponding to a volume element $dx_{1}\, dx_{2}\cdots dx_{n}$ should be $n$.
(with $dV_{E}$ corresponding to 3 dimensions). So the dimension function should
be additive where the volume element is multiplicative; this is the fundamental
property of a logarithm. So a natural choice for the effective spatial
dimension $d_{s}$ is
\begin{equation}
d_{s}=\frac{\ln\int_{0}^{\Lambda}dV'}{\ln\Lambda}.
\end{equation}
Evaluating this gives us
\begin{equation}
\label{eq-dswithLambda}
d_{s}=\frac{\ln\int_{0}^{\Lambda}dV_{E}}{\ln\Lambda}+2\frac{\ln\sqrt{g_{tt}}}
{\ln\Lambda}=3+2\frac{\ln\sqrt{g_{tt}}}{\ln\Lambda}.
\end{equation}

Equation~(\ref{eq-dswithLambda}) has some problems. The most striking one is
that $\Lambda$ appears to be a dimensional quantity, which would make $\frac{
\ln\sqrt{g_{tt}}}{\ln\Lambda}$ ambiguous. The obvious solution is that the
coordinates must be nondimensionalized, to make $\Lambda$ dimensionless.
However, it is not at all obvious how to nondimensionalize the coordinates.
Fortunately, we do not need to deal with that question directly. Regardless
of the coordinates' dimensions, $dV_{E}$ will always contribute three
dimensions to $d_{s}$. To analyze the last term in (\ref{eq-dswithLambda}),
we introduce the natural condition that the angular factors in the volume
element can not contribute any less than zero dimensions each.  That is,
\begin{equation}
\frac{\ln\sqrt{g_{tt}}}{\ln\Lambda}\geq-1.
\end{equation}
Since $g_{tt}<1$ and the logarithm is strictly increasing, this condition may
be rewritten as
\begin{equation}
\Lambda^{-1}\leq\sqrt{g_{tt}}.
\end{equation}
In classical General Relativity, $g_{tt}(r_{S})=0$, but in the
model~\cite{ref-chapline}, $g_{tt}$ drops to a nonzero minimum value
\begin{equation}
g_{tt}^{min}\sim 1-\frac{r_{S}}{r_{S}+\delta}\approx\frac{\delta}{r_{S}},
\end{equation}
where $\delta$ is a small length that characterizes how the classical
Schwarzschild singularity is cut off by the underlying quantum system. We
expect $\delta$ to be related to the Planck length. Thus we estimate $\Lambda
\sim\sqrt{r_{S}/\delta}$. Close to the event horizon, at $r=r_{S}+\Delta$,
where $\delta\ll\Delta\ll r_{S}$ and $g_{tt}\sim\Delta/r_{S}$, the effective
dimension of the system is
\begin{equation}
d_{s}=1+2\left(1-\frac{\ln\Delta/r_{S}}{\ln\delta/r_{S}}\right)=1+2\left(\frac
{\ln\delta/\Delta}{\ln\delta/r_{S}}\right).
\end{equation}
So we will look at the problem of Quantum Electrodynamics in $(1+\epsilon)+1$
dimensions, where
\begin{equation}
\epsilon=2\left(\frac{\ln\delta/\Delta}{\ln\delta/r_{S}}\right).
\end{equation}

We shall only consider the contribution to the photon self-energy from massless
particles. This should be a good approximation near the event horizon. Photons
coming in from spatial infinity are highly blueshifted at $r=r_{S}+\Delta$, so
the momentum scale in the photon propagator is large compared to any invariant
momentum scale (such as the electron mass). The same argument may also be
phrased in different terms. The energy of a comoving electron of mass $m_{e}$
is $\sqrt{g_{tt}}\, m_{e}$, so near the event horizon, the apparent electron
mass becomes small. So it is reasonable to consider massless particles.

The one-loop photon self-energy due to a single species of charged massless
fermions is
\begin{equation}
\label{eq-selfewithoutd}
i\Pi^{\mu\nu}(q)=2\,{\rm tr\, I}\,(q^{2}\eta^{\mu\nu}-q^{\mu}q^{\nu})
\int_{0}^{1}\! dx\int_{k}\frac{x(1-x)}{[k^{2}+q^{2}x(1-x)]^{2}}.
\end{equation}
Here, ${\rm I}$ is the unit matrix in spinor space, and the the Minkowski
metric is denoted by $\eta^{\mu\nu}$, to avoid confusion with the GR metric
$g^{\mu\nu}$. Since $d\rightarrow4$ as $r\rightarrow\infty$, the Dirac matrices
should be four-dimensional. Equation~(\ref{eq-selfewithoutd}), derived in the
appendix, gives, upon a $d$-dimensional $k$-integration,
\begin{equation}
\label{eq-selfenergy}
i\Pi^{\mu\nu}(q)= -i\,{\rm tr\, I}\,(q^{2}\eta^{\mu\nu}-q^{\mu}q^{\nu})\frac
{e^{2}}{(-q^{2})^{2-d/2}}
\frac{1}{(4\sqrt{\pi})^{d-1}}\frac{(1-d/2)\pi}{\sin\frac{\pi d}{2}}\frac{1}
{\Gamma(\frac{d}{2}+\frac{1}{2})}.
\end{equation}

There is a subtlety in the use of (\ref{eq-selfenergy}). In dimensional
regularization, it is usual to reduce all aspects of the problem to $d$
dimensions. In our case, only the loop integral is $d$-dimensional. There are
still four Dirac matrices, and the photon remains a four-component field.
However, so long as the external photon momentum $q$ lies in the
$d$-dimensional subspace, equation~(\ref{eq-selfenergy}) remains valid, and
the metric $\eta^{\mu\nu}$ is $d$-dimensional.

The $d=4$ and $d=2$ cases of equation~(\ref{eq-selfenergy}) are well
understood. Since we have $d=2+\epsilon$, we expand around $d=2$. Evaluating
equation~(\ref{eq-selfenergy}) with this value of $d$, we get
\begin{equation}
\label{eq-selfeatd}
i\Pi^{\mu\nu}(q) = -i(q^{2}\eta^{\mu\nu}-q^{\mu}q^{\nu})\frac{2e^{2}}{\pi}
\frac{1}{(-q^{2})^{1-\epsilon/2}}
\left\{\frac{\pi^{1/2-\epsilon/2}}{2^{1+2\epsilon}}\frac{\epsilon\pi/2}
{\sin\frac{\epsilon\pi}{2}}\frac{1}{\Gamma(\frac{3}{2}+\frac{\epsilon}{2})}
\right\}.
\end{equation}
The bracketed term in~(\ref{eq-selfeatd}) is unity at $\epsilon=0$. It is
purely real, so will only contribute higher-order corrections to the real and
imaginary parts of the self-energy.

At $d=2$, we get Schwinger's well-known result that the photon becomes
massive~\cite{ref-schwinger}. The self-energy is
\begin{equation}
\label{eq-selfeschwinger}
\left.i\Pi^{\mu\nu}(q)\right|_{d=2}=i\frac{2e^{2}}{\pi}\left(\eta^{\mu\nu}-
\frac{q^{\mu}q^{\nu}}{q^{2}}\right).
\end{equation}
(This differs by a factor of two from the usual result, because here we have
used four-dimensional Dirac matrices.) The residue of the pole at $q^{2}=0$
gives the photon mass $m_{\gamma}^{2}=\frac{2e^{2}}{\pi}$.

For $d=2+\epsilon$, the result is only slightly different. Instead of having
$\frac{1}{-q^{2}}$, we have $\frac{1}{(-q^{2})^{1-\epsilon/2}}$, which we
expand about $\epsilon=0$ to get
\begin{equation}
\label{eq-eexpansion}
\frac{1}{(-q^{2})^{1-\epsilon/2}}\approx\frac{1}{-q^{2}}\left[1+\frac{\epsilon}
{2}\ln(-q^{2})\right].
\end{equation}
As in equation~(\ref{eq-selfeschwinger}), there is a pole at $q^{2}=0$.
However, the residue is shifted by the second term in~(\ref{eq-eexpansion}). To
zeroth order in $\epsilon$, the pole in the propagator is shifted to $\frac
{2e^{2}}{\pi}$. To first order, the pole is further shifted to the value of
$[1+\frac{\epsilon}{2}\ln(-q^{2})]$ evaluated at the new pole location. This
shifts the pole to
\begin{eqnarray}
m_{\gamma}^{2} & = & \frac{2e^{2}}{\pi}\left[1+\frac{\epsilon}{2}\ln\left(
-\frac{2e^{2}}{\pi}\right)\right]\nonumber\\
& = & \frac{2e^{2}}{\pi}\left[1+\frac{\epsilon}{2}\ln(-1)+\frac{\epsilon}{2}\ln
\left(\frac{2e^{2}}{\pi}\right)\right]\nonumber\\
\label{eq-photmass}
& \approx & \frac{2e^{2}}{\pi}\left(1-i\frac{\pi\epsilon}{2}\right).
\end{eqnarray}
Equation~(\ref{eq-photmass}) is correct to lowest order in $\epsilon$ in both
its real and imaginary parts. The sign of the imaginary part has been chosen so
that photons decay rather than appear.

We must now turn to the question of how to interpret
equation~(\ref{eq-photmass}). By expanding around $d=2$, we have introduced
a number of two-dimensional conventions. The $e^{2}$ appearing in
(\ref{eq-photmass}) is the two-dimensional value of the electromagnetic
coupling. In two dimensions, $e$ has mass dimension one, so $\frac{2e^{2}}
{\pi}$ is indeed a mass squared. We must relate the $e$ in (\ref{eq-photmass})
(which we will henceforth refer to as $e_{2}$) to the four-dimensional electron
charge $e_{4}$.

We may relate the two charges by comparing the actions in two and four
dimensions. In four dimensions, the electromagnetic Lagrangian density is
\begin{equation}
{\cal L}_{4}=-\frac{1}{4e_{4}^{2}}F_{\mu\nu}F^{\mu\nu}.
\end{equation}
Then the action is
\begin{equation}
S_{4}=\int dt\, r^{2}dr\, d\Omega\,{\cal L}_{4}.
\end{equation}
The action $S_{2}$ derived from the two-dimensional La\-gran\-gi\-an ${\cal L}
_{2}$ should the same as $S_{4}$, up to a constant factor $C$. So we have
\begin{equation}
\label{eq-legrangians}
\int dt\, dr {\cal L}_{2}=-C\int dt\, r^{2}dr\, d\Omega\frac{1}{4e_{4}^{2}}
F_{\mu\nu}F^{\mu\nu}.
\end{equation}
We must perform the angular integrals on the right-hand side of
(\ref{eq-legrangians}) to determine ${\cal L}_{2}$. This means doing an
integral over the submanifold orthogonal to the $t$-$r$ subspace. This
orthogonal submanifold is a sphere, and the integral over it will depend upon
the radius at which the integral is evaluated. We are interested in in radii
$r\approx r_{S}$ (which is the only region where the integration over angles is
justified). If $F^{\mu\nu}$ is spherically symmetric, the angular integral
gives $4\pi$, and we can set $r=r_{S}$ to get
\begin{equation}
\int dt\, dr\, {\cal L}_{2}=-4\pi r_{S}^{2}C\int dt\, dr\,\frac{1}{4e_{4}^{2}}
F_{\mu\nu}F^{\mu\nu}.
\end{equation}
We can now read off the value of ${\cal L}_{2}$,
\begin{equation}
{\cal L}_{2}=-\frac{1}{4}\frac{4\pi r_{S}^{2}C}{e_{4}^{2}}F_{\mu\nu}F^{\mu\nu},
\end{equation}
so that the two-dimensional charge is
\begin{equation}
\label{eq-e2}
e_{2}^{2}=\frac{1}{4\pi r_{S}^{2}C}e_{4}^{2}.
\end{equation}
The constant $C$ includes differences in how the field operators are normalized
in two and four dimensions. So although the precise numerical relation between
$e_{2}$ and $e_{4}$ has not been determined, the dependence of $e_{2}$ on
$r_{S}$ is unambiguous.

In deriving equation~(\ref{eq-e2}), we assumed that the field configuration was
spherically symmetric. We can also evaluate the angular integral for more
general field configurations, although this adds additional ambiguities. If the
field $F^{\mu\nu}$ is in an $l>0$, $m=0$ multipole state, the integral
becomes
\begin{equation}
\label{eq-lneq0}
\int dt\, dr\, {\cal L}_{2}=-C\int dt\, r^{2}dr\, d\Omega\, P_{l}(\cos\theta)
^{2}\frac{1}{4e_{4}^{2}}F_{\mu\nu}F^{\mu\nu}.
\end{equation}
(The additional angular fields coming from derivatives of $P_{l}(\cos\theta)$
are suppressed by $\frac{1}{r_{S}}$ and have been dropped.) Since the maximum
value of $P_{l}(\cos\theta)$ is $P_{l}(1)=1$, the $F^{\mu\nu}$ appearing in
equation~(\ref{eq-lneq0}) is the maximum value of the field over all angles. It
is consistent with our earlier identification of the two- and four-dimensional
fields to identify this $F^{\mu\nu}$ with the field appearing in ${\cal L}_{2}$
although other identifications could also be consistent). Evaluating the
integral then gives us
\begin{equation}
\label{eq-e2lneq0}
e_{2}^{2}=\frac{2l+1}{4\pi r_{S}^{2}C}e_{4}^{2}.
\end{equation}
A similar calculation may be done for $m\neq0$, but the result (with these
conventions) depends on $m$ explicitly, not merely on $l$. Despite this
problem,~(\ref{eq-e2lneq0}) remains a good candidate for an $m$-independent
multipole field mass.

We must also address the question of how to interpret the imaginary part of
(\ref{eq-photmass}). To help with the interpretation, we shall use an analogy
to a much simpler dimensional reduction problem---the electromagnetic field in
a rectangular waveguide~\cite{ref-bekefi}. This simple problem in classical
electrodynamics has many similarities to the QED problem under consideration.

Consider a rectangular waveguide with metal walls. The waveguide has dimensions
$a$ in the $x$-direction and $b$ in the $y$-direction. (We will presume that
$a$ and $b$ are comparable in magnitude.) The waves propagate freely in the
$z$-direction. The boundary conditions on this system restrict the wavevector
of the electromagnetic field in the interior to be
\begin{equation}
{\bf k}=\frac{\pi n_{x}}{a}\hat{{\bf x}}+\frac{\pi n_{y}}{b}\hat{{\bf y}}+k_{z}
\hat{{\bf z}}.
\end{equation}
The numbers $n_{x}$ and $n_{y}$ are positive integers; at least one of $n_{x}$
and $n_{y}$ must be nonvanishing for fields to exist. The frequency $\omega=
|{\bf k}|$ satisfies
\begin{equation}
\label{eq-omega}
\omega^{2}=k_{z}^{2}+\pi^{2}\left(\frac{n_{x}^{2}}{a^{2}}+\frac{n_{y}^{2}}
{b^{2}}\right).
\end{equation}
Since propagation only occurs along the $z$-axis, it is natural to look at this
system in the $t$-$z$ subspace, where the wavevector is simply $k_{z}$. Then
(\ref{eq-omega}) looks like the energy-momentum relation for a relativistic
particle of mass $m_{wg}^{2}=\pi^{2}(\frac{n_{x}^{2}}{a^{2}}+\frac{n_{y}^{2}}
{b^{2}})$.

So in $1+1$ dimensions, a photon in a waveguide acquires an effective mass. The
scale of this mass is $a^{-1}$, where $a$ is the characteristic size of the
system in the neglected dimensions. This is the same scaling we found
previously. In equation~(\ref{eq-photmass}), the scale of the photon mass
was $r_{S}^{-1}$, and $r_{S}$ is the length scale of the event horizon in the
angular directions. According to (\ref{eq-e2lneq0}), the black hole system
actually has a whole hierarchy of photon masses. The waveguide also exhibits
this property; different $n_{x}$ and $n_{y}$ values give different values of
$m_{wg}^{2}$. (These results are similar to those found in Kaluza-Klein
theories, although the higher modes are not strongly suppressed here.)

The waveguide system also exhibits another important property---decay. Through
interactions in the $x$- and $y$-dimensions, a photon can disappear from the
interior of the waveguide. This can occur in a variety of ways, depending on
the regime. We mention the two simplest regimes and discuss the interpretation
of decays in these regimes. At low frequencies, $\omega\ll\nu$, where $\nu$ is
the collision frequency for electrons in the metal walls, the magnetic field
drives surface currents which dissipate energy through resistive heating. This
leads to a low-frequency energy loss
\begin{equation}
\Gamma\equiv\frac{\langle P\rangle}{\langle U\rangle}\propto\frac{1}{\sqrt
{a^{3}\sigma}},
\end{equation}
where $\sigma$ is the conductivity of the metal walls. The behavior is
different at higher frequencies, $\nu\ll\omega<\omega_{p}$, where $\omega_{p}$
is the plasma frequency, related to the electron density $n_{e}$ by $\omega_{p}
^{2}=\frac{n_{e}e^{2}}{m_{e}}$. In this regime, the electromagnetic field is
exponentially damped in the walls, but photons can tunnel through the walls and
escape from the waveguide. However, the decay rate does not have a simple
dependence on $a$ and $\omega_{p}$.

The decay rate in the waveguide depends strongly on the regime, but in each
regime, the decay rate depends primarily on the length $a$ and some other
length parameter. In the regimes outlined above, the length parameters are
provided by $\sigma$ and $\omega_{p}$. In the black hole model, the imaginary
part of $m_{\gamma}^{2}$ is governed by $\epsilon$, which depends on the
inverse mass scale $r_{S}$, as well as on the lengths $\delta$ and $\Delta$.

The two decay regimes outlined above possess very different decay processes.
The first regime is dissipative. Through interactions, photons in the waveguide
decay into something else---thermal excitations of the metal walls. In the
second regime, the photon does not actually decay. Instead, it escapes from the
waveguide and the corresponding $1+1$-dimensional subspace. Despite their
differences, these processes would each contribute an imaginary part to $m_{wg}
^{2}$.

Analogously, the imaginary term in equation~(\ref{eq-photmass}) could have two
different origins. The photons could be decaying into other particles, as is
suggested in~\cite{ref-chapline}. Alternatively, the effective decay rate in
(\ref{eq-photmass}) could correspond to photons being scattered out of the
$1+1$ dimensional $t$-$r$ subspace into states with large angular momenta. Our
simple treatment does not allow us to distinguish between the two. However,
either process would be novel---an effect caused by the finite minimum of $g
_{tt}$ and controlled in magnitude by $\delta$.

Although it is not directly relevant to the problem at hand, it is worth
mentioning one further suggestive aspect of the waveguide analogy. In $2+1$
dimensions, the photon self-energy (\ref{eq-selfenergy}) is still singular at
$q^{2}=0$, but that singularity is weaker than a simple pole in
$q^{2}$~\cite{ref-jackiw}. This introduces ambiguity as to whether or not the
photon has mass in $2+1$ dimensions. The waveguide analogy of $2+1$-dimensional
QED is the propagation of the electromagnetic field between two parallel plates
of separation $a$ in the $x$-direction. However, in this case, the wavevector
in the $x$-direction, $\frac{\pi n_{x}}{a}$, is allowed to vanish. So a photon
in this system may or may not behave as if it were massive.

\section*{Acknowledgments}

This work is supported in part by funds provided by the U. S. Department of
Energy (D.O.E.) under cooperative research agreement DE-FC02-94ER40818.

\appendix

\section*{Appendix: Photon self-energy integral}

To derive~(\ref{eq-selfewithoutd}) in a dimension-independent manner, we begin
with the integral
\begin{equation}
i\Pi^{\mu\nu}(q)=-e^{2}\int_{k}{\rm tr}\frac{\gamma^{\mu}(\!\not\! k+\!\not\! 
q)\gamma^{\nu}\!\not\! k}{(k+q)^{2}k^{2}}.
\end{equation}
Introducing a Feynman parameter $x$, setting $M^{2}=-x(1-x)q^{2}$, shifting
the integration variable $k\rightarrow k-xq$, and dropping all odd-$k$ terms
leaves
\begin{equation}
i\Pi^{\mu\nu}=-e^{2}\int_{0}^{1}\!dx\int_{k}{\rm tr}\frac{\gamma^{\mu}\!\!\not
\! k\gamma^{\nu}\!\!\not\! k-x(1-x)\gamma^{\mu}\!\not\! q\gamma^{\nu}\!\not\!
q}{(k^{2}-M^{2})^{2}}.
\end{equation}
Evaluating the traces for $D$-dimensional Dirac matrices and a $d$-dimensional
integration over $k$ gives
\begin{eqnarray}
{\rm tr}(\gamma^{\mu}\!\!\not\! k\gamma^{\nu}\!\!\not\! k) & = & {\rm tr}(-
\gamma^{\mu}\gamma^{\nu}k^{2}+2\gamma^{\mu}\!\!\not\! kk^{\nu})\nonumber\\
& = & {\rm tr}\left(-\gamma^{\mu}\gamma^{\nu}k^{2}+\frac{2}{d}\gamma^{\mu}
\gamma^{\nu}k^{2}\right)\nonumber\\
\label{eq-knumerator}
& = & \left(\frac{2}{d}-1\right)D\eta^{\mu\nu}k^{2}\\
{\rm tr}(\gamma^{\mu}\!\not\! q\gamma^{\nu}\!\not\! q) & = & -D\eta^{\mu\nu}q
^{2}+2Dq^{\mu}q^{\nu}
\label{eq-qnumerator}
\end{eqnarray}

The integrand with the $k^{2}$ from~(\ref{eq-knumerator}) may be transformed
to resemble the rest of the integrand according to
\begin{equation}
\label{eq-ksplit}
\int_{k}\frac{k^{2}}{(k^{2}-M^{2})^{2}}\left(\frac{2}{d}-1\right)=\int
_{k}\frac{1}{k^{2}-M^{2}}\left(\frac{2}{d}-1\right)
+\int_{k}\frac{M^{2}}{(k^{2}-M^{2})^{2}}\left(\frac{2}{d}-1\right).
\nonumber
\end{equation}
The first term of~(\ref{eq-ksplit}) may be Wick-rotated to Euclidean space and
evaluated using integration by parts, to give
\begin{eqnarray}
\int_{k}\frac{1}{k^{2}-M^{2}}\left(\frac{2}{d}-1\right) & = & i\int\frac{d
\Omega_{d}k^{d-1}dk}{-k^{2}-M^{2}}\left(\frac{2-d}{d}\right)\nonumber\\
& = & \frac{i}{d}\int\frac{d\Omega_{d}dk\, k^{2}}{k^{2}+M^{2}}(d-2)k^{d-3}
\nonumber\\
& = & \frac{-i}{n}\int\! d\Omega_{d}dk\,k^{d-2}\frac{d}{dk}\frac{k^{2}}{k^2+M
^{2}}\nonumber\\
& = & -\frac{2}{d}\int_{k}\frac{M^{2}}{(k^{2}+M^{2})^{2}}.
\end{eqnarray}
Adding this to the other term in~(\ref{eq-ksplit}) gives
\begin{equation}
\label{eq-kfinal}
\int_{k}\frac{k^{2}}{(k^{2}-M^{2})^{2}}\left(\frac{2}{d}-1\right)=-\int_{k}
\frac{M^{2}}{(k^{2}-M^{2})^{2}}.
\end{equation}
Combining~(\ref{eq-kfinal}) with the integrals having
numerator~(\ref{eq-qnumerator}) gives the self-energy result,
equation~(\ref{eq-selfewithoutd}).

\end{document}